\input epsf.sty
\documentclass[12pt]{article}

\begin{document}        
\pagestyle{empty}
\renewcommand{\thefootnote}{\fnsymbol{footnote}}

\begin{flushright}
{\small
SLAC--PUB--8169\\
May 1999\\}
\end{flushright}
 
\vspace{.8cm}

\begin{center}
{\bf\large   
A preliminary measurement of the gluon splitting rate into 
$b\bar{b}$ \\ \vskip 0.2cm
pairs in hadronic $Z^0$ decays
\footnote{Work supported by
Department of Energy contract  DE--AC03--76SF00515 (SLAC).}}

\vspace{1cm}
{\bf
Toshinori Abe} \\
\vskip 0.2cm
{\it
Stanford Linear Accelerator Center, Stanford University,
Stanford, CA  94309
} \\
\vskip 0.6cm
{\bf
Representing the SLD Collaboration$^{**}$
} \\
\vskip 0.2cm
\medskip
\end{center}
 
\vfill

\begin{center}
{\bf\large   
Abstract }
\end{center}
We present a measurement of the rate of gluon splitting into bottom quarks, 
$g \to b\bar{b}$, 
in hadronic $Z^0$ decays collected by SLD from 1996 to 1998.
The analysis was performed by looking for secondary bottom production in
4-jet events of any primary flavor.  A topological vertex mass technique was
used to tag the two jets with the smallest angle between them as $b/\bar{b}$.
We obtained a rate of $g \to b\bar{b}$ per hadronic event to be
$(3.07 \pm 0.71 {\rm (stat.)} \pm 0.66 {\rm (syst.)})\times 10^{-3}$ 
(preliminary).
\vfill

\begin{center} 
{\it Presented at the American Physical Society (APS) Meeting of 
the Division of Particles and Fields (DPF 99), 5-9 January 1999, 
University of California, Los Angeles} 
\end{center}

\newpage

 
 
%
\pagestyle{plain}

\section{Introduction}               

The process of the splitting of a gluon into a heavy-quark pair is 
one of the elementary processes in QCD but
is poorly known, both theoretically and experimentally.

The rate $g_{b\bar{b}}$ is defined as the fraction of hadronic
events in which a gluon splits into a $b\bar{b}$ pair,
$e^+e^- \to q\bar{q}g \to q\bar{q}b\bar{b}$.
The value of $g_{b\bar{b}}$ is an infrared finite quantity, 
because the $b$-quark mass provides a natural cutoff, 
hence it can be safely computed in the framework of perturbative QCD
\cite{Miller:1998ig}.
However the rate is sensitive to the $\alpha_S$ parameter 
and to the $b$-quark mass, 
which results in a substantial theoretical uncertainty
in the calculation of $g_{b\bar{b}}$.
The limited accuracy of the $g_{b\bar{b}}$ prediction is one of the
main sources of uncertainty in the measurement of the partial decay
width  
$R_{b}=\Gamma (Z^0 \to b\bar{b})/ \Gamma (Z^0 \to q\bar{q})$ \cite{RbSLD,RbLEP}.
In addition, about $50$\% 
of the B hadrons produced at the Tevatron are due to
the gluon splitting process,
and a larger fraction is expected to contribute at the LHC.
A better knowledge of this process can improve theoretical
predictions of heavy-flavor production at such colliders.

This measurement is difficult experimentally.
The cross section of $g \to b\bar{b}$ is very small even at $Z^0$ energies,
since the gluon must have sufficient mass to produce the bottom-quark pair.
There are huge backgrounds from $Z^0 \to b\bar{b}$
whose magnitude is about a hundred times larger than the 
$Z^0 \to q\bar{q}g \to q\bar{q}b\bar{b}$ process.
Moreover the B hadrons from $g \to b\bar{b}$ have relatively low energy and  
short flight distance and are more difficult to distinguish using 
standard vertexing.
So far, the only three measurements of $g_{b\bar{b}}$
have been reported, by DELPHI and ALEPH \cite{GBBLEP}.

Here we present a new measurement of $g_{b\bar{b}}$
based on a 400k $Z^0$-decay data
sample taken in 1996-98 at the Stanford Linear Collider (SLC),
with the SLC Large Detector (SLD). 
In this period, $Z^0$ decays were collected 
with an upgraded vertex detector,
wider acceptance and better impact parameter resolution, 
thus improving considerably the $b$-tagging performance.

\section{The SLD Detector}

A full description of the SLD and its performance have been described 
in detail elsewhere \cite{SLD}.
Only the details most relevant to this analysis are mentioned here.

SLD is well-suited for the measurement of $g\to b\bar{b}$ due to
two unique features.
The first is that the SLC, 
the only linear collider in the world, 
provides a very small and stable beam spot.
The SLC interaction point was reconstructed from tracks in sets of 
approximately thirty sequential hadronic $Z^0$ decays with an uncertainty 
of only $5\mu$m transverse to the beam axis and $32\mu$m 
(for $b\bar{b}$ events) along the beam axis.
Second is the upgraded vertex detector (VXD3) \cite{Abe:1997bu},
a pixel-based CCD vertex detector.
VXD3 consists of 3 layers with 300M pixels and
each layer is only $0.36\%$ of a radiation length thick.
The measured $r\phi$ ($rz$) track impact-parameter resolution approaches
$11\mu$m ($23\mu$m) for high momentum tracks, while multiple scattering 
contributions are $40\mu{\rm m}/(p_{\perp}\sin^{3/2}\theta)$ in
both projections ($z$ is the coordinate parallel to the beam axis and
$p_{\perp}$ is the momentum in GeV/c perpendicular to the beam line).
With these features, topological
vertex finding gives excellent $b$-tagging efficiency and purity.
In particular, the efficiency is good even at low B-meson energies,
which is especially important for detecting $g \to b\bar{b}$.

\section{Flavor Tagging \label{SEC:BTAG}} 
 
Topologically reconstructed secondary vertices \cite{ZVNIM}
are used by many analyses at the SLD for heavy-quark tagging.
To reconstruct the secondary vertices, the space points 
where track density functions overlap are found in 3-dimensions.
Only the vertices that are significantly displaced from the primary
vertex (PV) are considered to be possible B- or D-hadron
decay vertices.
The mass of the secondary vertex
is calculated using the tracks that
are associated with the vertex. 
Since the heavy-hadron decays are frequently accompanied by neutral particles,
the reconstructed mass is corrected to account for this fact.
By using kinematic information from
the vertex flight path and
the momentum sum of the tracks associated with the secondary vertex,
we calculate the $P_T$-corrected mass $M_{P_T}$ 
by adding a minimum amount of missing momentum to the invariant mass,
as follows:
$$M_{P_T} = \sqrt{{M^2}_{VTX} + {P_T}^2} + |P_T|.$$
Here $M_{VTX}$ is the invariant mass of the tracks associated with 
the reconstructed secondary vertex and $P_T$ is the transverse momentum
of the charged tracks with respect to the B-flight direction.
In this correction, vertexing resolution as well as the PV resolution
are crucial. 
Due to the small and stable interaction point at the SLC and 
the excellent vertexing resolution from the SLD CCD Vertex detector, 
this technique has so far only been successfully applied at the SLD.

\section{Monte Carlo and data Samples}

The measurement uses 400k events collected from 1996 to 1998 with the 
requirement that the VXD3 was fully operational.

For the purpose of estimating the efficiency and purity of 
the $g\to b\bar{b}$ selection procedure, 
we made use of a detailed Monte-Carlo simulation of the detector.
The JETSET 7.4 \cite{JETSET} event generator was used, 
with parameter values tuned to hadronic $e^+e^-$ annihilation 
data \cite{LUNDTUNE}, 
combined with a simulation of B hadron decays  
tuned to $\Upsilon(4S)$ data \cite{SLDSIM} 
and a simulation of the SLD based on GEANT 3.21 \cite{GEANT}.
Inclusive distributions of single-particle and event-topology observables
in hadronic events were found to be well described by the
simulations \cite{SLDALPHAS}.
Uncertainties in the simulation
were taken into account in the systematic errors (Section~\ref{SEC:SYS}).

Monte-Carlo events are reweighted to take into account current 
estimates for gluon splitting into heavy-quark pairs \cite{GBBLEP,GCCLEP}.
The JETSET at SLD predicts $g_{b\bar{b}}=0.14\%$ and $g_{c\bar{c}}=1.36\%$,
and we reweighted them so that 
$g_{b\bar{b}}=0.273\%$ and $g_{c\bar{c}}=2.58\%$.
A Monte-Carlo production of about 1200k $Z\to q\bar{q}$ events,
1000k $Z\to b\bar{b}$ events and 480k $Z\to c\bar{c}$ events are used
in order to better evaluate the efficiencies.

Besides the signal events, hereafter called B, two categories of background
events exist:
\begin{itemize}
\item	Events which do not contain any gluon splitting into heavy flavor 
	at all, hereafter called Q events; and
\item	Events in which a gluon splits to a charm quark pair, named C events.
\end{itemize}

\section{Event Selection}

The two B hadrons coming from the gluon tend to be produced in a particular 
topological configuration, 
which allows one to discriminate the signal from background.
We select $g \to b\bar{b}$ events as follows:
\begin{itemize}
\item	Require 4 jets in the events;
\item	Require $b$ tags in two jets selected in a particular configuration; 
	and
\item	Apply additional topological selections to improve the
	signal/background ratio.
\end{itemize}

Jets are formed with energy-flow particles, using the Durham jet-finding
algorithm \cite{Catani:1991hj} with $y_{cut}=0.008$, 
chosen to minimize the statistical error.
The 4-jet fractions for the B, C and Q events predicted by the simulation are
about $32\%$, $18\%$ and $5.3\%$, respectively.
The overall 4-jet rate in the data is $(5.976\pm0.044)\%$, 
where the error is statistical only.
In the Monte-Carlo simulation the rate is $(5.678\pm0.002\pm0.068)\%$ 
where the first error is statistical and the second is due to the 
uncertainty in the simulation of heavy-quark-hadron physics.
The two jets forming the smallest angle in the event are considered as
candidates for originating from the gluon splitting process $g\to b\bar{b}$.
The selected jets are labeled as jet 1 and jet 2, where jet 1 is more
energetic than jet 2.
The other two jets in the event are labeled as jets 3 and 4,
where jet 3 is more energetic than jet 4.

Jets containing B-hadron decay products are then searched for by making
use of the information coming from the vertex detector,
using the topological vertex method.
We require jet 1 and jet 2 to each have a secondary vertex.
No tag is applied to jet 3 and jet 4.
After topological vertexing, about 300 events are selected.
The selection efficiency for $g \to b\bar{b}$ is expected from Monte Carlo
to be $6.6\%$
while the signal/background ratio is $1/5$.
$67\%$ of the background comes from $Z \to b\bar{b}$ events, 
$21\%$ from $g \to c\bar{c}$ events and remaining $12\%$ from 
$Z \to q\bar{q}$ ($q \ne b$) events.

\begin{figure}[h]	
\centerline{\epsfxsize 2.5 truein \epsfbox{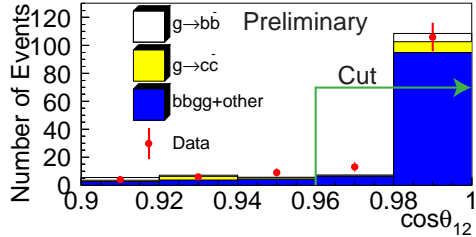}}   
\vskip -.2 cm
\caption[]{
\label{COS12}
\small	Angular distribution between vertex axes in jet 1 and jet 2
($0.9<\cos\theta_{12}$).
Points indicate data, open box signal,
hatched boxes are backgrounds.
}
\end{figure}

In order to improve the signal/background ratio, 
we use topological information.
Firstly, many $b\bar{b}$ background have one $b$-jet splitting
into 2 jets so that the two found vertices are from different decay products
from the same B decay.
The two vertex axes tend to be collinear.
Figure \ref{COS12} shows the angular distribution between vertex axes
in jet 1 and jet 2.
Half of the $b\bar{b}$ background peaks at $\cos\theta_{12} \sim 1$.
In order to remove $b\bar{b}$ events, 
we require $-0.2 < \cos\theta_{12} < 0.96$.
%

\begin{figure}[h]	
\centerline{\epsfxsize 2.5 truein \epsfbox{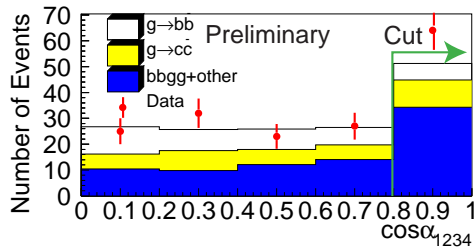}}
\vskip -.2 cm
\caption[]{
\label{COS1234}
\small	 The distribution of the cosine of angle between 
the plane $\Pi_{12}$ formed by jets 1 and 2 and the plane $\Pi_{34}$
formed by jets 3 and 4,
for data (points) and Monte Carlo (histogram).	
We reject $|\cos\alpha_{1234}|>0.8$}
\end{figure}

Secondly, the variable $|\cos\alpha_{1234}|$, where $\alpha_{1234}$
is the angle between the plane $\Pi_{12}$ formed by jets 1 and 2
and the plane $\Pi_{34}$ by jets 3 and 4,
is used to suppress the $b\bar{b}$ background.
Figure \ref{COS1234} shows the distribution of $|\cos\alpha_{1234}|$.
This variable is similar to the Bengtsson-Zerwas angle \cite{BZANGLE},
and is useful to separate $g \to b\bar{b}$ events
because the radiated virtual gluon in the process $Z^0\to q\bar{q}g$
is polarized in the plane of the three-parton event, 
and this is reflected in its subsequent splitting, by strongly
favoring $g \to q\bar{q}$ emission out of this plane.
Events with $|\cos\alpha_{1234}|>0.8$ are rejected.
%

Thirdly, the $b$ jets coming from a gluon tend to have lower energy
than the other two jets in the event.
We require the jet-energy sum of jet 1 and jet 2 to be smaller than
36 GeV.

\begin{figure}[h]	
\centerline{\epsfxsize 2.5 truein \epsfbox{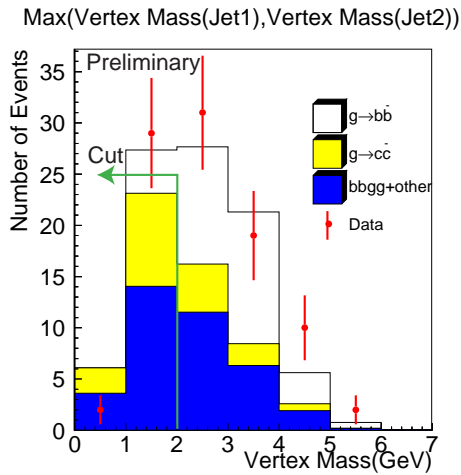}}   
\vskip -.2 cm
\caption[]{
\label{VMASS2}
\small	Maximum $P_T$-corrected mass distribution between jet 1 and jet 2 
	after jet-energy-sum cut.
	Points indicate data, open box signal,
	hatched boxes are backgrounds.}
\end{figure}

Finally,
$c$ jets have lower $P_{T}$-corrected mass than $b$ jets.
Figure \ref{VMASS2} shows the greater of the $P_T$-corrected mass 
determined for jet 1 and jet 2 after the above cuts.
Many $g\to c\bar{c}$ events are in below 2.0 GeV.
Hence we require maximum $P_T$-corrected mass to be greater 
than 2.0 GeV to remove $g\to c\bar{c}$ events.

\section{Result}
After requiring all the above mentioned cuts, 
62 events are selected in the data.
Background events are estimated to be $27.6$ using Monte Carlo,
where $63\%$ of the background comes from $Z \to b\bar{b}$ events, 
$27\%$ from $g \to c\bar{c}$ events
and the remaining $10\%$ from $Z \to q\bar{q}$ $(q \ne b)$ events.
Table \ref{SELEFF} shows the tagging efficiencies for the three categories
of events, 
where the errors are statistical only.
From these efficiencies and the fraction of events selected in the data
$f_{d}=(2.14\pm0.27)\times 10^{-4}$, 
the value of $g_{b\bar{b}}$ can be extracted as:
\begin{equation}
g_{b\bar{b}}=\frac{f_{d} - ( 1 - g_{c\bar{c}} )\epsilon_{Q} - g_{c\bar{c}}
	\epsilon_{C}}{\epsilon_{B} - \epsilon_{Q}}.
\end{equation}

The measured value of the gluon splitting rate into $b\bar{b}$ pairs is
\begin{equation}
g_{b\bar{b}}=(3.07 \pm 0.71) \times 10^{-3},
\end{equation}
where the error is statistical only.

\begin{table}
\centerline{
\begin{tabular}{cc} 
  \hline\hline
  Source & Efficiency ($\%$) \\  
  \hline
  B & $3.86 \pm 0.52$ \\
  C & $0.10 \pm 0.02$ \\
  Q & $0.73 \pm 0.05$ \\
  \hline\hline
\end{tabular}
}
\caption{
\label{SELEFF}
 Efficiencies after all cuts for the three categories.
 Errors are statistical only.}
\end{table}

\section{Systematic Error \label{SEC:SYS}}

The efficiencies for the three event categories are evaluated by Monte-Carlo
simulation.
The limitations of the simulation in estimating these efficiencies lead to 
an uncertainty on the result.
The error due to the limited Monte-Carlo statistics in the efficiency
evaluation is $\Delta g_{b\bar{b}} = \pm 0.44 \times 10^{-3}$.
This uncertainty comes mainly from the efficiency to tag Q events.

A large fraction of events remaining after the selection cuts
contain $b$ and $c$ hadrons.
The uncertainty in the knowledge of the physical processes in the simulation
of heavy-flavor production and decays constitutes a source of systematic
error.
All the physical simulation parameters are varied within their allowed 
experimental ranges.
In particular, the $b$ and $c$ hadron lifetimes as well as production rates
are varied, following the latest recommendations of the LEP Heavy
Flavour Working Group \cite{LEPHF}.
The uncertainties are summarized in Table \ref{SYSERR}.

The simulation of the signal events is based on the JETSET parton shower Monte
Carlo, which is in good agreement with the theoretical predictions
\cite{Miller:1998ig}.
In order to estimate the uncertainty on this assumption, 
we have produced 10,000 $g \to b\bar{b}$ events using 
GRC4F \cite{Fujimoto:1997wj}
at the generator level.
The signal tagging efficiency mainly depends on the energy of the gluon
splitting into $b\bar{b}$.
This efficiency function, computed with JETSET, is reweighted by 
the ratio of GRC4F to JETSET initial distributions to obtain the average
efficiency.
A systematic error of $\pm 0.09 \times 10^{-3}$ is estimated from
the difference in $\epsilon_{{\rm B}}$ from the two Monte-Carlo models.

The dependence of the B efficiency on the $b$-quark mass has also been 
investigated at the generator level.
Events are generated using the GRC4F Monte Carlo,
which is based on a matrix element calculation including $b$-quark masses.
The variation of the B efficiency is computed as done for JETSET, 
using the GRC4F spectrum for $b$-quark masses from $4.7$ and $5.3$ GeV/c$^2$.
The uncertainty is estimated to be $0.06 \times 10^{-3}$.

The uncertainty in the ratio of the $g \to c\bar{c}$ background events,
$\Delta g_{c\bar{c}} = \pm 0.40{\rm \%}$, gives the error 
$\Delta g_{b\bar{b}} = \pm 0.09 \times 10^{-3}$.

There is about $5\%$ discrepancy of 4-jet rate 
between data and Monte Carlo in our $y_{cut}$.
The uncertainty due to the discrepancy is estimated by increasing
background events in the Monte Carlo to be
$\Delta g_{b\bar{b}} = \pm 0.14 \times 10^{-3}$.

Charged Monte-Carlo tracks used by the topological vertex tag are
smeared and tossed to better reproduce distribution of data.
Uncertainties in the efficiencies due to this smearing and tossing 
are assessed by evaluating the Monte-Carlo efficiencies without 
the smearing and tossing algorithm.
The difference in the $g_{b\bar{b}}$ result is taken as systematic error.
The errors on $g_{b\bar{b}}$ due to the tracking resolution and efficiency 
are then estimated as 
$\Delta g_{b\bar{b}} = \pm 0.26 \times 10^{-3}$
and $= \pm 0.29 \times 10^{-3}$, respectively.

Table \ref{SYSERR} summarizes the different sources of systematic error
on $g_{b\bar{b}}$, and the total systematic error is estimated to be
$0.66 \times 10^{-3}$.
%
\begin{table}
\centerline{
\begin{tabular}{lc} 
  \hline\hline
  Source & $\Delta g_{b\bar{b}}$ $(10^{-3})$ \\
  \hline
  Monte Carlo statistics			& $\pm 0.44$ \\
  $b$ hadron lifetimes				& $\pm 0.01$ \\
  $b$ hadron production				& $\pm 0.07$ \\
  $b$ hadron fragmentation			& $\pm 0.12$ \\
  $b$ hadron charged multiplicities		& $\pm 0.11$ \\
  $c$ hadron lifetimes				& $\pm 0.01$ \\
  $c$ hadron production				& $\pm 0.03$ \\
  $c$ hadron charged multiplicities		& $\pm 0.08$ \\
  Energy distribution of $g\to b\bar{b}$	& $\pm 0.08$ \\
  $b$ quark mass					& $\pm 0.06$ \\
  $g_{c\bar{c}}$				& $\pm 0.09$ \\
  4-jet rate discrepancy			& $\pm 0.14$ \\
  IP resolution				& $\pm 0.09$ \\
  Track resolution				& $\pm 0.26$ \\
  Tracking efficiency				& $\pm 0.29$ \\
  \hline
  Total (Preliminary)				& $\pm 0.66$ \\
  \hline\hline
\end{tabular}
}
\caption{
\label{SYSERR}
  Systematic uncertainties on $g_{b\bar{b}}$.
}
\end{table}

\section{Summary}

A measurement of the gluon splitting rate to a $b\bar{b}$ pair in hadronic
$Z^0$ decays collected by SLD has been presented.
Excellent SLC and VXD3 performance provides advantages not only for
$b$-tag efficiency but also for topological selections.
The result is
$$
g_{b\bar{b}}=(3.07\pm0.71{\rm (stat.)}\pm0.66{\rm (syst.)})\times10^{-3}
\rm{(preliminary)}.
$$
where the first error is statistical and the second includes all 
systematic effects.

\def\pl{Phys.\ Lett.\ }


\newpage

\section*{$^{**}$List of Authors} 


%
%
%
\begin{center}
\def\iADEL{$^{(1)}$}
\def\iAOMORI{$^{(2)}$}
\def\iBOLO{$^{(3)}$}
\def\iBRI{$^{(4)}$}
\def\iBRUN{$^{(5)}$}
\def\iBU{$^{(6)}$}
\def\iCINC{$^{(7)}$}
\def\iCOLO{$^{(8)}$}
\def\iCOLU{$^{(9)}$}
\def\iCSU{$^{(10)}$}
\def\iFERR{$^{(11)}$}
\def\iFRAS{$^{(12)}$}
\def\iILLI{$^{(13)}$}
\def\iJHU{$^{(14)}$}
\def\iLBL{$^{(15)}$}
\def\iLTU{$^{(16)}$}
\def\iMASS{$^{(17)}$}
\def\iMISSI{$^{(18)}$}
\def\iMIT{$^{(19)}$}
\def\iMOSCOW{$^{(20)}$}
\def\iNAGO{$^{(21)}$}
\def\iOREG{$^{(22)}$}
\def\iOXF{$^{(23)}$}
\def\iPADO{$^{(24)}$}
\def\iPERU{$^{(25)}$}
\def\iPISA{$^{(26)}$}
\def\iRAL{$^{(27)}$}
\def\iRUTG{$^{(28)}$}
\def\iSLAC{$^{(29)}$}
\def\iSOGA{$^{(30)}$}
\def\iSOONG{$^{(31)}$}
\def\iTENN{$^{(32)}$}
\def\iTOHO{$^{(33)}$}
\def\iUCSB{$^{(34)}$}
\def\iUCSC{$^{(35)}$}
\def\iUVIC{$^{(36)}$}
\def\iVAND{$^{(37)}$}
\def\iWASH{$^{(38)}$}
\def\iWISC{$^{(39)}$}
\def\iYALE{$^{(40)}$}

  \baselineskip=.75\baselineskip  
\mbox{Kenji  Abe\unskip,\iNAGO}
\mbox{Koya Abe\unskip,\iTOHO}
\mbox{T. Abe\unskip,\iSLAC}
\mbox{I. Adam\unskip,\iSLAC}
\mbox{T.  Akagi\unskip,\iSLAC}
\mbox{N.J. Allen\unskip,\iBRUN}
\mbox{W.W. Ash\unskip,\iSLAC}
\mbox{D. Aston\unskip,\iSLAC}
\mbox{K.G. Baird\unskip,\iMASS}
\mbox{C. Baltay\unskip,\iYALE}
\mbox{H.R. Band\unskip,\iWISC}
\mbox{M.B. Barakat\unskip,\iLTU}
\mbox{O. Bardon\unskip,\iMIT}
\mbox{T.L. Barklow\unskip,\iSLAC}
\mbox{G.L. Bashindzhagyan\unskip,\iMOSCOW}
\mbox{J.M. Bauer\unskip,\iMISSI}
\mbox{G. Bellodi\unskip,\iOXF}
\mbox{R. Ben-David\unskip,\iYALE}
\mbox{A.C. Benvenuti\unskip,\iBOLO}
\mbox{G.M. Bilei\unskip,\iPERU}
\mbox{D. Bisello\unskip,\iPADO}
\mbox{G. Blaylock\unskip,\iMASS}
\mbox{J.R. Bogart\unskip,\iSLAC}
\mbox{G.R. Bower\unskip,\iSLAC}
\mbox{J.E. Brau\unskip,\iOREG}
\mbox{M. Breidenbach\unskip,\iSLAC}
\mbox{W.M. Bugg\unskip,\iTENN}
\mbox{D. Burke\unskip,\iSLAC}
\mbox{T.H. Burnett\unskip,\iWASH}
\mbox{P.N. Burrows\unskip,\iOXF}
\mbox{A. Calcaterra\unskip,\iFRAS}
\mbox{D. Calloway\unskip,\iSLAC}
\mbox{B. Camanzi\unskip,\iFERR}
\mbox{M. Carpinelli\unskip,\iPISA}
\mbox{R. Cassell\unskip,\iSLAC}
\mbox{R. Castaldi\unskip,\iPISA}
\mbox{A. Castro\unskip,\iPADO}
\mbox{M. Cavalli-Sforza\unskip,\iUCSC}
\mbox{A. Chou\unskip,\iSLAC}
\mbox{E. Church\unskip,\iWASH}
\mbox{H.O. Cohn\unskip,\iTENN}
\mbox{J.A. Coller\unskip,\iBU}
\mbox{M.R. Convery\unskip,\iSLAC}
\mbox{V. Cook\unskip,\iWASH}
\mbox{R.F. Cowan\unskip,\iMIT}
\mbox{D.G. Coyne\unskip,\iUCSC}
\mbox{G. Crawford\unskip,\iSLAC}
\mbox{C.J.S. Damerell\unskip,\iRAL}
\mbox{M.N. Danielson\unskip,\iCOLO}
\mbox{M. Daoudi\unskip,\iSLAC}
\mbox{N. de Groot\unskip,\iBRI}
\mbox{R. Dell'Orso\unskip,\iPERU}
\mbox{P.J. Dervan\unskip,\iBRUN}
\mbox{R. de Sangro\unskip,\iFRAS}
\mbox{M. Dima\unskip,\iCSU}
\mbox{A. D'Oliveira\unskip,\iCINC}
\mbox{D.N. Dong\unskip,\iMIT}
\mbox{M. Doser\unskip,\iSLAC}
\mbox{R. Dubois\unskip,\iSLAC}
\mbox{B.I. Eisenstein\unskip,\iILLI}
\mbox{V. Eschenburg\unskip,\iMISSI}
\mbox{E. Etzion\unskip,\iWISC}
\mbox{S. Fahey\unskip,\iCOLO}
\mbox{D. Falciai\unskip,\iFRAS}
\mbox{C. Fan\unskip,\iCOLO}
\mbox{J.P. Fernandez\unskip,\iUCSC}
\mbox{M.J. Fero\unskip,\iMIT}
\mbox{K. Flood\unskip,\iMASS}
\mbox{R. Frey\unskip,\iOREG}
\mbox{J. Gifford\unskip,\iUVIC}
\mbox{T. Gillman\unskip,\iRAL}
\mbox{G. Gladding\unskip,\iILLI}
\mbox{S. Gonzalez\unskip,\iMIT}
\mbox{E.R. Goodman\unskip,\iCOLO}
\mbox{E.L. Hart\unskip,\iTENN}
\mbox{J.L. Harton\unskip,\iCSU}
\mbox{A. Hasan\unskip,\iBRUN}
\mbox{K. Hasuko\unskip,\iTOHO}
\mbox{S.J. Hedges\unskip,\iBU}
\mbox{S.S. Hertzbach\unskip,\iMASS}
\mbox{M.D. Hildreth\unskip,\iSLAC}
\mbox{J. Huber\unskip,\iOREG}
\mbox{M.E. Huffer\unskip,\iSLAC}
\mbox{E.W. Hughes\unskip,\iSLAC}
\mbox{X. Huynh\unskip,\iSLAC}
\mbox{H. Hwang\unskip,\iOREG}
\mbox{M. Iwasaki\unskip,\iOREG}
\mbox{D.J. Jackson\unskip,\iRAL}
\mbox{P. Jacques\unskip,\iRUTG}
\mbox{J.A. Jaros\unskip,\iSLAC}
\mbox{Z.Y. Jiang\unskip,\iSLAC}
\mbox{A.S. Johnson\unskip,\iSLAC}
\mbox{J.R. Johnson\unskip,\iWISC}
\mbox{R.A. Johnson\unskip,\iCINC}
\mbox{T. Junk\unskip,\iSLAC}
\mbox{R. Kajikawa\unskip,\iNAGO}
\mbox{M. Kalelkar\unskip,\iRUTG}
\mbox{Y. Kamyshkov\unskip,\iTENN}
\mbox{H.J. Kang\unskip,\iRUTG}
\mbox{I. Karliner\unskip,\iILLI}
\mbox{H. Kawahara\unskip,\iSLAC}
\mbox{Y.D. Kim\unskip,\iSOGA}
\mbox{M.E. King\unskip,\iSLAC}
\mbox{R. King\unskip,\iSLAC}
\mbox{R.R. Kofler\unskip,\iMASS}
\mbox{N.M. Krishna\unskip,\iCOLO}
\mbox{R.S. Kroeger\unskip,\iMISSI}
\mbox{M. Langston\unskip,\iOREG}
\mbox{A. Lath\unskip,\iMIT}
\mbox{D.W.G. Leith\unskip,\iSLAC}
\mbox{V. Lia\unskip,\iMIT}
\mbox{C.Lin\unskip,\iMASS}
\mbox{M.X. Liu\unskip,\iYALE}
\mbox{X. Liu\unskip,\iUCSC}
\mbox{M. Loreti\unskip,\iPADO}
\mbox{A. Lu\unskip,\iUCSB}
\mbox{H.L. Lynch\unskip,\iSLAC}
\mbox{J. Ma\unskip,\iWASH}
\mbox{G. Mancinelli\unskip,\iRUTG}
\mbox{S. Manly\unskip,\iYALE}
\mbox{G. Mantovani\unskip,\iPERU}
\mbox{T.W. Markiewicz\unskip,\iSLAC}
\mbox{T. Maruyama\unskip,\iSLAC}
\mbox{H. Masuda\unskip,\iSLAC}
\mbox{E. Mazzucato\unskip,\iFERR}
\mbox{A.K. McKemey\unskip,\iBRUN}
\mbox{B.T. Meadows\unskip,\iCINC}
\mbox{G. Menegatti\unskip,\iFERR}
\mbox{R. Messner\unskip,\iSLAC}
\mbox{P.M. Mockett\unskip,\iWASH}
\mbox{K.C. Moffeit\unskip,\iSLAC}
\mbox{T.B. Moore\unskip,\iYALE}
\mbox{M.Morii\unskip,\iSLAC}
\mbox{D. Muller\unskip,\iSLAC}
\mbox{V. Murzin\unskip,\iMOSCOW}
\mbox{T. Nagamine\unskip,\iTOHO}
\mbox{S. Narita\unskip,\iTOHO}
\mbox{U. Nauenberg\unskip,\iCOLO}
\mbox{H. Neal\unskip,\iSLAC}
\mbox{M. Nussbaum\unskip,\iCINC}
\mbox{N. Oishi\unskip,\iNAGO}
\mbox{D. Onoprienko\unskip,\iTENN}
\mbox{L.S. Osborne\unskip,\iMIT}
\mbox{R.S. Panvini\unskip,\iVAND}
\mbox{C.H. Park\unskip,\iSOONG}
\mbox{T.J. Pavel\unskip,\iSLAC}
\mbox{I. Peruzzi\unskip,\iFRAS}
\mbox{M. Piccolo\unskip,\iFRAS}
\mbox{L. Piemontese\unskip,\iFERR}
\mbox{K.T. Pitts\unskip,\iOREG}
\mbox{R.J. Plano\unskip,\iRUTG}
\mbox{R. Prepost\unskip,\iWISC}
\mbox{C.Y. Prescott\unskip,\iSLAC}
\mbox{G.D. Punkar\unskip,\iSLAC}
\mbox{J. Quigley\unskip,\iMIT}
\mbox{B.N. Ratcliff\unskip,\iSLAC}
\mbox{T.W. Reeves\unskip,\iVAND}
\mbox{J. Reidy\unskip,\iMISSI}
\mbox{P.L. Reinertsen\unskip,\iUCSC}
\mbox{P.E. Rensing\unskip,\iSLAC}
\mbox{L.S. Rochester\unskip,\iSLAC}
\mbox{P.C. Rowson\unskip,\iCOLU}
\mbox{J.J. Russell\unskip,\iSLAC}
\mbox{O.H. Saxton\unskip,\iSLAC}
\mbox{T. Schalk\unskip,\iUCSC}
\mbox{R.H. Schindler\unskip,\iSLAC}
\mbox{B.A. Schumm\unskip,\iUCSC}
\mbox{J. Schwiening\unskip,\iSLAC}
\mbox{S. Sen\unskip,\iYALE}
\mbox{V.V. Serbo\unskip,\iSLAC}
\mbox{M.H. Shaevitz\unskip,\iCOLU}
\mbox{J.T. Shank\unskip,\iBU}
\mbox{G. Shapiro\unskip,\iLBL}
\mbox{D.J. Sherden\unskip,\iSLAC}
\mbox{K.D. Shmakov\unskip,\iTENN}
\mbox{C. Simopoulos\unskip,\iSLAC}
\mbox{N.B. Sinev\unskip,\iOREG}
\mbox{S.R. Smith\unskip,\iSLAC}
\mbox{M.B. Smy\unskip,\iCSU}
\mbox{J.A. Snyder\unskip,\iYALE}
\mbox{H. Staengle\unskip,\iCSU}
\mbox{A. Stahl\unskip,\iSLAC}
\mbox{P. Stamer\unskip,\iRUTG}
\mbox{H. Steiner\unskip,\iLBL}
\mbox{R. Steiner\unskip,\iADEL}
\mbox{M.G. Strauss\unskip,\iMASS}
\mbox{D. Su\unskip,\iSLAC}
\mbox{F. Suekane\unskip,\iTOHO}
\mbox{A. Sugiyama\unskip,\iNAGO}
\mbox{S. Suzuki\unskip,\iNAGO}
\mbox{M. Swartz\unskip,\iJHU}
\mbox{A. Szumilo\unskip,\iWASH}
\mbox{T. Takahashi\unskip,\iSLAC}
\mbox{F.E. Taylor\unskip,\iMIT}
\mbox{J. Thom\unskip,\iSLAC}
\mbox{E. Torrence\unskip,\iMIT}
\mbox{N.K. Toumbas\unskip,\iSLAC}
\mbox{T. Usher\unskip,\iSLAC}
\mbox{C. Vannini\unskip,\iPISA}
\mbox{J. Va'vra\unskip,\iSLAC}
\mbox{E. Vella\unskip,\iSLAC}
\mbox{J.P. Venuti\unskip,\iVAND}
\mbox{R. Verdier\unskip,\iMIT}
\mbox{P.G. Verdini\unskip,\iPISA}
\mbox{D.L. Wagner\unskip,\iCOLO}
\mbox{S.R. Wagner\unskip,\iSLAC}
\mbox{A.P. Waite\unskip,\iSLAC}
\mbox{S. Walston\unskip,\iOREG}
\mbox{J. Wang\unskip,\iSLAC}
\mbox{S.J. Watts\unskip,\iBRUN}
\mbox{A.W. Weidemann\unskip,\iTENN}
\mbox{E. R. Weiss\unskip,\iWASH}
\mbox{J.S. Whitaker\unskip,\iBU}
\mbox{S.L. White\unskip,\iTENN}
\mbox{F.J. Wickens\unskip,\iRAL}
\mbox{B. Williams\unskip,\iCOLO}
\mbox{D.C. Williams\unskip,\iMIT}
\mbox{S.H. Williams\unskip,\iSLAC}
\mbox{S. Willocq\unskip,\iMASS}
\mbox{R.J. Wilson\unskip,\iCSU}
\mbox{W.J. Wisniewski\unskip,\iSLAC}
\mbox{J. L. Wittlin\unskip,\iMASS}
\mbox{M. Woods\unskip,\iSLAC}
\mbox{G.B. Word\unskip,\iVAND}
\mbox{T.R. Wright\unskip,\iWISC}
\mbox{J. Wyss\unskip,\iPADO}
\mbox{R.K. Yamamoto\unskip,\iMIT}
\mbox{J.M. Yamartino\unskip,\iMIT}
\mbox{X. Yang\unskip,\iOREG}
\mbox{J. Yashima\unskip,\iTOHO}
\mbox{S.J. Yellin\unskip,\iUCSB}
\mbox{C.C. Young\unskip,\iSLAC}
\mbox{H. Yuta\unskip,\iAOMORI}
\mbox{G. Zapalac\unskip,\iWISC}
\mbox{R.W. Zdarko\unskip,\iSLAC}
\mbox{J. Zhou\unskip.\iOREG}

\it
  \vskip \baselineskip                   
  \centerline{(The SLD Collaboration)}   
  \vskip \baselineskip        
  \baselineskip=.75\baselineskip   
\iADEL
  Adelphi University, Garden City, New York 11530, \break
\iAOMORI
  Aomori University, Aomori , 030 Japan, \break
\iBOLO
  INFN Sezione di Bologna, I-40126, Bologna, Italy, \break
\iBRI
  University of Bristol, Bristol, U.K., \break
\iBRUN
  Brunel University, Uxbridge, Middlesex, UB8 3PH United Kingdom, \break
\iBU
  Boston University, Boston, Massachusetts 02215, \break
\iCINC
  University of Cincinnati, Cincinnati, Ohio 45221, \break
\iCOLO
  University of Colorado, Boulder, Colorado 80309, \break
\iCOLU
  Columbia University, New York, New York 10533, \break
\iCSU
  Colorado State University, Ft. Collins, Colorado 80523, \break
\iFERR
  INFN Sezione di Ferrara and Universita di Ferrara, I-44100 Ferrara, Italy, \break
\iFRAS
  INFN Lab. Nazionali di Frascati, I-00044 Frascati, Italy, \break
\iILLI
  University of Illinois, Urbana, Illinois 61801, \break
\iJHU
  Johns Hopkins University,  Baltimore, Maryland 21218-2686, \break
\iLBL
  Lawrence Berkeley Laboratory, University of California, Berkeley, California 94720, \break
\iLTU
  Louisiana Technical University, Ruston,Louisiana 71272, \break
\iMASS
  University of Massachusetts, Amherst, Massachusetts 01003, \break
\iMISSI
  University of Mississippi, University, Mississippi 38677, \break
\iMIT
  Massachusetts Institute of Technology, Cambridge, Massachusetts 02139, \break
\iMOSCOW
  Institute of Nuclear Physics, Moscow State University, 119899, Moscow Russia, \break
\iNAGO
  Nagoya University, Chikusa-ku, Nagoya, 464 Japan, \break
\iOREG
  University of Oregon, Eugene, Oregon 97403, \break
\iOXF
  Oxford University, Oxford, OX1 3RH, United Kingdom, \break
\iPADO
  INFN Sezione di Padova and Universita di Padova I-35100, Padova, Italy, \break
\iPERU
  INFN Sezione di Perugia and Universita di Perugia, I-06100 Perugia, Italy, \break
\iPISA
  INFN Sezione di Pisa and Universita di Pisa, I-56010 Pisa, Italy, \break
\iRAL
  Rutherford Appleton Laboratory, Chilton, Didcot, Oxon OX11 0QX United Kingdom, \break
\iRUTG
  Rutgers University, Piscataway, New Jersey 08855, \break
\iSLAC
  Stanford Linear Accelerator Center, Stanford University, Stanford, California 94309, \break
\iSOGA
  Sogang University, Seoul, Korea, \break
\iSOONG
  Soongsil University, Seoul, Korea 156-743, \break
\iTENN
  University of Tennessee, Knoxville, Tennessee 37996, \break
\iTOHO
  Tohoku University, Sendai 980, Japan, \break
\iUCSB
  University of California at Santa Barbara, Santa Barbara, California 93106, \break
\iUCSC
  University of California at Santa Cruz, Santa Cruz, California 95064, \break
\iUVIC
  University of Victoria, Victoria, British Columbia, Canada V8W 3P6, \break
\iVAND
  Vanderbilt University, Nashville,Tennessee 37235, \break
\iWASH
  University of Washington, Seattle, Washington 98105, \break
\iWISC
  University of Wisconsin, Madison,Wisconsin 53706, \break
\iYALE
  Yale University, New Haven, Connecticut 06511. \break

\rm
%

\end{center}

\end{document}